# Quantization of the electromagnetic field, entropy of an ideal monoatomic gas, and the birth of Bose-Einstein statistics

Masud Mansuripur

James C. Wyant College of Optical Sciences, The University of Arizona, Tucson, Arizona, USA

**Abstract**. In 1924, Einstein received a short manuscript in the mail from the Indian physicist S.N. Bose. He quickly translated Bose's manuscript to German and submitted it to *Zeitschrift für Physik*. Within a few weeks, Einstein presented his own findings (using a generalization of Bose's counting method) to a session of the *Prussian Academy of Sciences*. Whereas Bose had suggested a new counting method for the quanta of the electromagnetic field — one that yielded Planck's blackbody radiation formula — Einstein applied Bose's method to an ideal monoatomic gas. Shortly afterward, Einstein presented to the *Academy* a follow-up paper in which he described the Bose-Einstein condensation for the first time. The present paper describes some of the fascinating issues that Einstein struggled with as he attempted to unify the quantum-statistical properties of matter with those of the electromagnetic field.

**1. Introduction**. When Bose's brief manuscript on a novel method of counting the phase-space configurations of light quanta (later dubbed photons) in thermal equilibrium inside a closed box was rejected by *Philosophical Magazine*, he sent it (along with a letter, dated June 4, 1924) to Einstein in Germany. Einstein immediately recognized the significance of Bose's insight and proceeded to translate the manuscript into German and submit it to *Zeitschrift für Physik*. Believing in the unity of nature and that what is true for the quanta of the electromagnetic field must also be true for material particles, Einstein extended Bose's method and applied it to the molecules (or atoms) of an ideal gas. He thus discovered the inadequacy of the conventional Maxwell-Boltzmann statistics, as well as that of the traditional estimates of the entropy of an ideal gas. In short order, Einstein predicted the existence of a new phase of matter — what has come to be known as the Bose-Einstein condensate — that could coexist with the rest of the gas (in a saturated phase) at sufficiently low temperatures. The crucial elements of the Bose-Einstein statistics are that (i) there is no limit on the number of identical particles (bosons) that could co-exist in a given state, and (ii) an exchange of two identical bosons, each in a different state, does *not* give rise to a new configuration.

The goal of the present paper is to examine the original arguments of Bose and Einstein, give a detailed derivation of some important findings that were only sketched in those early papers, and comment on several points that were hinted at but not duly elaborated at the time. In the next section, we reproduce Bose's original argument that treats photons in thermal equilibrium inside an otherwise empty box as individual particles distributed among well-defined cells in the position-momentum phase space — a treatment that culminates in Planck's blackbody radiation formula. Also derived in this section are expressions for the energy, the entropy, and the pressure of the photon gas in thermal equilibrium at temperature $T$ within a closed box of volume $V$. Section 3 is devoted to an alternative, albeit equivalent, treatment of Bose's distribution of photons among the various cells in the phase space. The method of counting the number of distinct phase-space configurations described in Sec.3 was proposed by Einstein as a (mathematically) simpler approach to arriving at Bose's original results. Einstein also proposed a "mutually statistically independent" treatment of the particles in their localization, which resulted in Wien's displacement law (rather than in Planck's distribution) for the photon gas. Einstein's method of counting the phase-space configurations in accordance with his statistically-independent hypothesis is elaborated in Sec.4.

The main concern of Einstein's two-part paper, of course, was an extension of Bose's counting method to the atoms (or molecules) of an ideal gas. In Sec.5, we describe how Einstein arrived at his conclusion that an ideal monoatomic gas (in thermal equilibrium within a closed box), if cooled down below a critical temperature, must undergo a transition to a two-phase state. This is where Einstein predicts that, at sufficiently low temperatures, a zero-energy, zero-entropy phase should separate from the rest of the (monoatomic) gas to form a condensate. Section 5 also examines the pressure and the specific heat of an ideal gas in accordance with the Bose-Einstein statistics. In Sec.6, we show that Einstein's alternative (but equivalent) method of counting the configurations of atoms in the phase space yields identical results to those obtained in Sec.5. His "statistically-independent" hypothesis, however, leads to the classical Maxwell-Boltzmann distribution and a violation of Nernst's theorem, as elaborated in Sec.7. Einstein concluded that the Maxwell-Boltzmann statistics and also the Sackur-Tetrode expression of the entropy of an ideal gas (both emerging from his "statistically independent" treatment) are valid only approximately and at higher temperatures, whereas the counting of the phase-space configurations in accordance with Bose's method leads to correct statistics and, therefore, correct physics all the way down to zero temperature. The paper closes with a few concluding remarks and an amusing anecdote in Sec.8. Two topics that were discussed by Einstein but may be considered as digressions from the main themes of his two-part paper, are treated in detail in Appendices A and B at the end of the paper.

**2. Bose's derivation of Planck's law based on the light quantum hypothesis.**[1] Consider a closed box of volume $V$ filled with thermal radiation of total electromagnetic energy $\mathcal{E}$. Inside the box, individual photons of frequency $\nu_s$ have energy $h\nu_s$ and linear momentum $h\nu_s/c$ along their forward propagation direction; consequently, $|\boldsymbol{p}|^2 = p_x^2 + p_y^2 + p_z^2 = (h\nu_s/c)^2$. We take the frequencies $\nu_s$ to be discrete samples from a continuous spectrum of positive frequencies, with the integer index $s$ ranging from zero to infinity. We denote by $\mathrm{d}\nu_s$ the infinitesimal width of the narrow frequency interval within which $\nu_s$ is the center frequency. The volume of the six-dimensional phase space associated with the frequency range $\mathrm{d}\nu_s$ is given by

$$\int \mathrm{d}x\mathrm{d}y\mathrm{d}z\mathrm{d}p_x\mathrm{d}p_y\mathrm{d}p_z = V \times \overbrace{4\pi(h\nu_s/c)^2}^{\text{surface area of sphere}} \times \overbrace{(h/c)\mathrm{d}\nu_s}^{\text{spherical shell thickness}} = 4\pi V(h^3\nu_s^2/c^3)\mathrm{d}\nu_s. \quad (1)$$

The units of $\mathrm{d}x\mathrm{d}p_x$ (and, similarly, those of $\mathrm{d}y\mathrm{d}p_y$ and $\mathrm{d}z\mathrm{d}p_z$) are $\mathrm{kg}\cdot\mathrm{m}^2/\mathrm{s} = \mathrm{joule}\cdot\mathrm{s}$, which coincide with the units of Planck's constant $h$. This is also the minimum uncertainty in the position-momentum product for each of the three coordinates of the $xyz$-space in accordance with Heisenberg's uncertainty principle. Postulating that the phase space is a contiguous collection of elementary cells of volume $h^3$, and accounting for the fact that individual photons can be in one of two allowed polarization states (say, right- and left-circular[2]), the total number of elementary cells inhabiting the accessible phase space in the immediate vicinity of $\nu_s$ is seen to be[1]

$$N_s = 8\pi V(\nu_s^2/c^3)\mathrm{d}\nu_s. \quad (2)$$

Note that $N_s$ is dimensionless, as it must be, considering that the elementary cell volume $h^3$ has the dimensions of the left-hand side of Eq.(1). Among these $N_s$ cells, let there be $n_{0,s}$ cells that contain no photons, $n_{1,s}$ cells that contain a single photon, $n_{2,s}$ cells that contain two photons, and so on.[1] Clearly,

$$N_s = \textstyle\sum_{r=0}^{\infty} n_{r,s}. \quad (3)$$

The overall energy content of our radiation-filled box is thus given by

$$\mathcal{E} = \sum_{s=0}^{\infty} \sum_{r=0}^{\infty} r n_{r,s} h\nu_s. \tag{4}$$

The number of distinct arrangements of the cells associated with $\mathrm{d}\nu_s$ is $N_s!/(n_{0,s}!\, n_{1,s}!\, n_{2,s}! \cdots)$. Taken across the entire range of frequencies $\nu_0, \nu_1, \nu_2, \cdots$, the number of distinct arrangements of the entirety of the phase-space cells is found to be

$$w = \prod_{s=0}^{\infty} [N_s! \prod_{r=0}^{\infty} (n_{r,s}!)^{-1}]. \tag{5}$$

The goal now is to maximize $w$ subject to the constraints of Eqs.(3) and (4). Recalling the Stirling[3] upper and lower bounds on $n!$, namely, $(n/e)^n \sqrt{n} e^{7/8} < n! < (n/e)^n \sqrt{n} e$, and taking $N_s$ and $n_{r,s}$ to be sufficiently large numbers to justify the approximation $n! \cong (n/e)^n$, we write†

$$\ln(w) \cong \sum_{s=0}^{\infty} (N_s \ln N_s) - \sum_{s=0}^{\infty} \sum_{r=0}^{\infty} [n_{r,s} \ln(n_{r,s})]. \tag{6}$$

We recognize that $N_s$, whose value is given by Eq.(2), is independent of the variables $n_{r,s}$ of the function $\ln(w)$ appearing in Eq.(6). Applying the method of Lagrange multipliers[3] to maximize $\ln(w)$ subject to the constraints of Eqs.(3) and (4) requires the unconstrained optimization of the following composite function over the space of its independent variables $n_{r,s}$:

Lagrange multipliers

$$\sum_{s=0}^{\infty} \sum_{r=0}^{\infty} [n_{r,s} \ln(n_{r,s})] + \sum_{s=0}^{\infty} \lambda_s \left(\sum_{r=0}^{\infty} n_{r,s}\right) + \zeta \sum_{s=0}^{\infty} \sum_{r=0}^{\infty} r n_{r,s} h\nu_s. \tag{7}$$

Equating to zero the derivatives of the above function with respect of each and every variable $n_{r,s}$, we arrive at ‡

$$[\ln(n_{r,s}) + 1] + \lambda_s + \zeta r h\nu_s = 0 \qquad \rightarrow \qquad n_{r,s} = e^{-(1+\lambda_s)} e^{-\zeta r h\nu_s}. \tag{8}$$

To satisfy the first constraint imposed on $n_{r,s}$, we substitute from Eq.(8) into Eq.(3), arriving at

$$N_s = \sum_{r=0}^{\infty} n_{r,s} = e^{-(1+\lambda_s)} \sum_{r=0}^{\infty} e^{-\zeta r h\nu_s} = e^{-(1+\lambda_s)}/(1 - e^{-\zeta h\nu_s}). \quad \leftarrow \boxed{\sum_{r=0}^{\infty} x^r = 1/(1-x)} \tag{9}$$

This yields a solution for $e^{-(1+\lambda_s)}$, which goes back into Eq.(8) to yield

$$n_{r,s} = 8\pi V(\nu_s^2/c^3)(1 - e^{-\zeta h\nu_s}) e^{-\zeta r h\nu_s} \mathrm{d}\nu_s. \quad \leftarrow \boxed{\text{see Eq.(2)}} \tag{10}$$

As for the second constraint given by Eq.(4), substitution from Eq.(10) leads to§

$$\begin{aligned}
\mathcal{E} = \sum_{s=0}^{\infty} \sum_{r=0}^{\infty} r n_{r,s} h\nu_s &= (8\pi hV/c^3) \sum_{s=0}^{\infty} \nu_s^3 (1 - e^{-\zeta h\nu_s}) \left(\sum_{r=0}^{\infty} r e^{-\zeta r h\nu_s}\right) \mathrm{d}\nu_s \\
&= (8\pi hV/c^3) \sum_{s=0}^{\infty} \nu_s^3\, e^{-\zeta h\nu_s} \mathrm{d}\nu_s/(1 - e^{-\zeta h\nu_s}) \\
&= (8\pi hV/c^3) \sum_{s=0}^{\infty} \nu_s^3 (e^{\zeta h\nu_s} - 1)^{-1} \mathrm{d}\nu_s \\
\boxed{\text{Gradshteyn \& Ryzhik}^4\ \textbf{3.411-1};\ \text{also}\ \textbf{4.262-2}} \rightarrow &= \frac{8\pi V}{(ch)^3 \zeta^4} \int_0^{\infty} \frac{x^3}{e^x - 1} \mathrm{d}x = \frac{8\pi^5 V}{15 (ch)^3 \zeta^4} = \frac{\pi^2 V}{15 c^3 \hbar^3 \zeta^4}. \quad \leftarrow \boxed{\hbar = h/2\pi}
\end{aligned} \tag{11}$$

The above equation yields the value of the Lagrange multiplier $\zeta$ in terms of the energy $\mathcal{E}$ trapped inside the box — or, as is often referred to, the cavity. It is desirable, however, to express the energy

---

† In the approximate version of Eq.(5), the term $\exp(-N_s)$ is cancelled out by $\exp(\sum_{r=0}^{\infty} n_{r,s})$; see Eq.(3).

‡ The derivative of the function $n_{r,s} \ln(n_{r,s})$ with respect to $n_{r,s}$ (treated as a continuous variable) is $\ln(n_{r,s}) + 1$. Had we *not* applied the Stirling approximation to $n_{r,s}!$, this term would have been $\ln(n_{r,s}!)$, whose (discrete) derivative is $\ln(n_{r,s} + 1)$ on the right-hand side, and $\ln(n_{r,s})$ on the left-hand side.

§ For $|x| < 1$, we have: $\sum_{r=0}^{\infty} x^r = 1/(1-x) \quad \rightarrow \quad \sum_{r=0}^{\infty} r x^{r-1} = 1/(1-x)^2 \quad \rightarrow \quad \sum_{r=0}^{\infty} r x^r = x/(1-x)^2$.

$\mathcal{E}$ of the photon gas in terms of the cavity temperature $T$. To this end, we proceed to evaluate the entropy[1] of the photon gas from Eq.(6) with the aid of Eqs.(8) and (9), as follows:

$$S = k_B \ln(w) \cong k_B\{\textstyle\sum_{s=0}^{\infty}(N_s \ln N_s) - \sum_{s=0}^{\infty}\sum_{r=0}^{\infty}[n_{r,s}\ln(n_{r,s})]\} \quad \leftarrow \boxed{k_B \text{ is Boltzmann's constant}}$$

$$= k_B[\textstyle\sum_{s=0}^{\infty}(N_s \ln N_s) + \sum_{s=0}^{\infty}\sum_{r=0}^{\infty} n_{r,s}(1+\lambda_s+\zeta r h\nu_s)] \quad \leftarrow \boxed{\text{see Eq.(8)}}$$

$$= k_B[\textstyle\sum_{s=0}^{\infty}(N_s \ln N_s) + \sum_{s=0}^{\infty} N_s(1+\lambda_s) + \sum_{s=0}^{\infty}\sum_{r=0}^{\infty} n_{r,s}(\zeta r h\nu_s)]$$

$$\boxed{\text{from Eq.(9): } 1+\lambda_s = -\ln[N_s(1-e^{-\zeta h\nu_s})]}$$

$$= k_B\{\textstyle\sum_{s=0}^{\infty}(N_s \ln N_s) - \sum_{s=0}^{\infty} N_s \ln[N_s(1-e^{-\zeta h\nu_s})] + \zeta\sum_{s=0}^{\infty} h\nu_s \sum_{r=0}^{\infty}(r n_{r,s})\}$$

$$= k_B\{\zeta\mathcal{E} - \textstyle\sum_{s=0}^{\infty}[N_s\ln(1-e^{-\zeta h\nu_s})]\} \quad \leftarrow \boxed{\ln(1-x) = -\sum_{n=1}^{\infty} x^n/n}$$

$$= k_B\{\zeta\mathcal{E} + (8\pi V/c^3)\textstyle\sum_{s=0}^{\infty}\nu_s^2 \mathrm{d}\nu_s \sum_{n=1}^{\infty} n^{-1}e^{-n\zeta h\nu_s}\} \quad \leftarrow \boxed{\text{see Eq.(2)}}$$

$$= k_B\left\{\zeta\mathcal{E} + (8\pi V/c^3)\textstyle\sum_{n=1}^{\infty} n^{-1}\int_{\nu_s=0}^{\infty}\nu_s^2 e^{-n\zeta h\nu_s}\mathrm{d}\nu_s\right\}$$

$$= k_B\{\zeta\mathcal{E} + (8\pi V/c^3)\textstyle\sum_{n=1}^{\infty} n^{-1}(n\zeta h)^{-3}\int_0^{\infty} x^2 e^{-x}\mathrm{d}x\} \quad \leftarrow \boxed{\text{integration by parts: } \int_0^{\infty} x^2e^{-x}\mathrm{d}x = 2}$$

$$= k_B\{\zeta\mathcal{E} + [16\pi V/(c\zeta h)^3]\textstyle\sum_{n=1}^{\infty} n^{-4}\} \quad \leftarrow \boxed{\sum_{n=1}^{\infty} n^{-4} = \pi^4/90}$$

$$= k_B\left[\zeta\mathcal{E} + \frac{8\pi^5 V}{45(c\zeta h)^3}\right] = k_B\left[\frac{\pi^2 V}{15(c\zeta\hbar)^3} + \frac{\pi^2 V}{45(c\zeta\hbar)^3}\right] = \frac{4\pi^2 k_B V}{45(c\zeta\hbar)^3}. \quad \leftarrow \boxed{\text{see Eq.(11); also } \hbar = h/2\pi} \tag{12}$$

Invoking the thermodynamic identity $\partial S/\partial\mathcal{E} = 1/T$, where $T$ is the absolute temperature of the photon gas in thermal equilibrium with the walls of its container (volume $V$ = constant),[5,6] we now obtain the value of the Lagrange multiplier $\zeta$ from Eqs.(11) and (12), as follows:

$$\frac{1}{T} = \frac{\partial S/\partial\zeta}{\partial\mathcal{E}/\partial\zeta} = \frac{-3(4\pi^2 k_B V)}{45(c\hbar)^3\zeta^4} \div \frac{-4(\pi^2 V)}{15(c\hbar)^3\zeta^5} = k_B\zeta \quad \rightarrow \quad \zeta = 1/(k_B T). \tag{13}$$

Substitution into Eq.(11) reveals the dependence of the total energy $\mathcal{E}$ on the temperature $T$ and volume $V$ of the cavity; that is,

$$\mathcal{E} = (8\pi h V/c^3)\int_0^{\infty}\nu_s^3(e^{h\nu_s/k_B T}-1)^{-1}\mathrm{d}\nu_s = \frac{\pi^2 V k_B^4 T^4}{15c^3\hbar^3}. \tag{14}$$

It is now possible to see how the entropy $S$ of Eq.(12) arises from the energy $\mathcal{E}$ of Eq.(14). Given that $\delta S = \delta Q/T$, where $\delta Q$ is the thermal energy (or heat) needed to raise the entropy by $\delta S$ at fixed $T$ and $V$, and since $\delta Q = \delta\mathcal{E} = 4\pi^2 V k_B^4 T^3\delta T/(15c^3\hbar^3)$, we will have

$$S = \int_0^T \delta Q/T = [4\pi^2 V k_B^4/(15c^3\hbar^3)]\int_0^T T^2\delta T = 4\pi^2 V k_B^4 T^3/(45c^3\hbar^3). \tag{15}$$

The above result is the same entropy as in Eq.(12) with $\zeta = 1/(k_B T)$ in accordance with Eq.(13). Note that both the energy $\mathcal{E}$ and the entropy $S$ of the photon gas approach zero as $T \rightarrow 0$, even though the approximation used in taking Eq.(5) to Eq.(6) may not be justifiable at very low temperatures. The pressure $p$ of the photon gas can be obtained from Eqs.(14) and (15) by forming the Helmholtz free energy[7] $F = \mathcal{E} - TS = -\pi^2 V k_B^4 T^4/(45c^3\hbar^3)$, then computing $p = -\partial F/\partial V$, which yields $p = \mathcal{E}/3V$. Planck's blackbody energy-density of the photon gas (per unit volume per unit frequency) appears in the penultimate line of Eq.(11) after substituting $1/(k_B T)$ for $\zeta$ as $(8\pi h\nu_s^3/c^3)/[\exp(h\nu_s/k_B T) - 1]$.

**3. Einstein's alternative (but equivalent) counting method**. Following the publication of Bose's original paper,[1] Einstein published a two-part paper[7,8] in which he extended Bose's treatment of the photon gas to a monoatomic gas consisting of $\mathbb{N}$ identical particles in thermal equilibrium with their container — a closed box of volume $V$. In the second part of his paper,[8] Einstein suggested an alternative (albeit equivalent) method of counting the number of distinct arrangements in which different numbers of identical particles could occupy the various cells of the phase-space. Einstein's alternative counting involves arrangements in which $N_s$ elementary cells inhabiting the phase space of photons in the frequency interval $[\nu_s, \nu_s + \mathrm{d}\nu_s)$ are occupied by a total of $M_s$ identical (i.e., indistinguishable) photons. As explained in Fig.1, the total number $w$ of distinct arrangements obtained via this method, which is equivalent to Bose's original counting method, can be written as

$$w = \prod_{s=0}^{\infty} (M_s + N_s - 1)!/[M_s!\,(N_s - 1)!]. \tag{16}$$

Invoking the Stirling[3] approximation $n! \cong (n/e)^n$, we write the natural logarithm of $w$ as follows:

$$\ln(w) \cong \sum_{s=0}^{\infty} [(M_s + N_s - 1)\ln(M_s + N_s - 1) - M_s \ln(M_s) - (N_s - 1)\ln(N_s - 1)]. \tag{17}$$

The last term on the right-hand side of Eq.(17) is independent of $M_s$; this is readily verified by noting that $N_s$ is given by Eq.(2). Maximizing the remaining part of $\ln(w)$ subject to the constraint that the overall energy of the photon gas must be $\mathcal{E} = \sum_{s=0}^{\infty} M_s h\nu_s$, requires consideration of the following composite function:

Lagrange multiplier

$$\sum_{s=0}^{\infty} [(M_s + N_s - 1)\ln(M_s + N_s - 1) - M_s \ln(M_s)] - \zeta\left(\sum_{s=0}^{\infty} M_s h\nu_s\right). \tag{18}$$

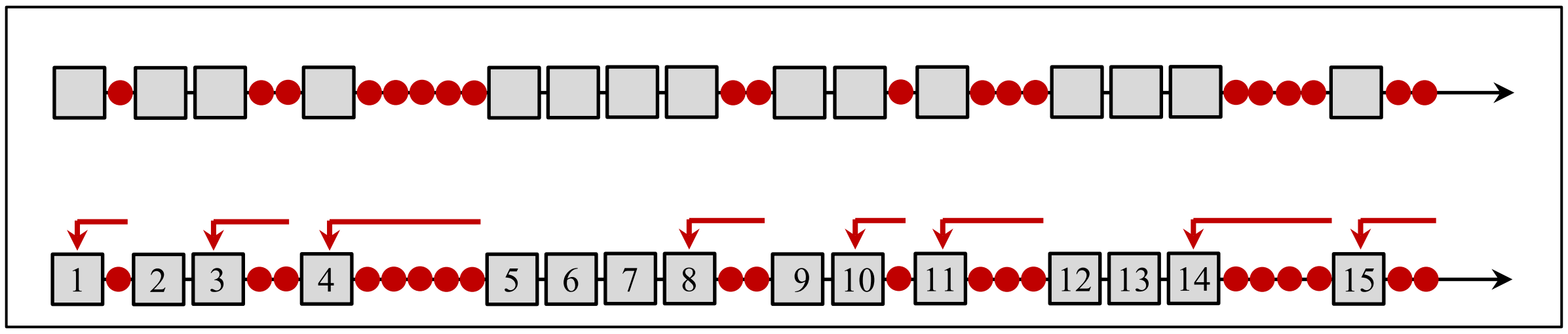

**Fig.1**. A set consisting of $M_s$ photons in the frequency interval $[\nu_s, \nu_s + \mathrm{d}\nu_s)$, depicted as red dots, and the corresponding $N_s$ available cells of the phase space, depicted as grey boxes (volume $= h^3$), is arranged randomly along a straight line, much like a mixed set of beads on a string. (Here, $M_s = 20$ and $N_s = 15$.) The first element on the left-hand side is required to be a grey box in all admissible arrangements. At first, we assume that the $M_s$ photons are distinct (i.e., distinguishable from each other), and also that the remaining $N_s - 1$ boxes are distinct. Thus, the number of different arrangements of the collection of red dots + grey boxes is $(M_s + N_s - 1)!$. Since the photons are, in principle, indistinguishable from one another, this overall number of arrangements must be divided by $M_s!$. We now take the grey boxes to also be indistinguishable, which requires a further dividing of the number of distinct arrangements by $(N_s - 1)!$. All in all, the total number of distinct arrangements of red dots + grey boxes is seen to be $(M_s + N_s - 1)!/[M_s!\,(N_s - 1)!]$. At this point, we identify the grey boxes as the sequentially-ordered cells of the phase space, and label them (from left to right) as cell numbers $1, 2, 3, \cdots, N_s$. Subsequently, the photons residing between adjacent pairs of cells are moved into the cells that are immediately to the left of each such group of photons; this is indicated by the arrows appearing above each group of photons. In this way, some of the cells remain empty (i.e., they will contain zero photons), some will have a single photon, a few will be occupied by two photons, and so on. The total number of distinct distributions of $M_s$ identical photons among $N_s$ sequentially-labeled cells is thus seen to be $(M_s + N_s - 1)!/[M_s!\,(N_s - 1)!]$.

The independent variables of the above function are $M_s$, whose optimal values can be found by equating to zero the derivatives of the function with respect to each and every $M_s$; this yields

$$\ln(M_s + N_s - 1) - \ln(M_s) - \zeta h\nu_s = 0 \qquad \rightarrow \qquad M_s = (N_s - 1)/(e^{\zeta h\nu_s} - 1). \qquad (19)$$

The value of the Lagrange multiplier $\zeta$ may now be determined by enforcing the constraint on the total energy of the photon gas — which is in thermal equilibrium with the walls of its container (volume $= V$). We will have [$N_s - 1 \cong N_s$] [see Eq.(2)]

$$\mathcal{E} = \textstyle\sum_{s=0}^{\infty} M_s h\nu_s \cong \sum_{s=0}^{\infty} N_s (e^{\zeta h\nu_s} - 1)^{-1} h\nu_s = (8\pi hV/c^3) \sum_{s=0}^{\infty} \nu_s^3 (e^{\zeta h\nu_s} - 1)^{-1} \mathrm{d}\nu_s$$

$$= (8\pi hV/c^3) \int_0^{\infty} \nu_s^3 (e^{\zeta h\nu_s} - 1)^{-1} \mathrm{d}\nu_s = \frac{\pi^2 V}{15 c^3 \hbar^3 \zeta^4}. \qquad (20)$$

[G&R **3.411**-1; $\hbar = h/2\pi$]

This equation contains the same expressions as found previously in Eq.(11) for the energy-density of blackbody radiation and the corresponding total energy $\mathcal{E}$ of the confined photon gas. The entropy of the photon gas may now be obtained from Eq.(17), as follows:

$$S = k_B \ln(w) \cong k_B \textstyle\sum_{s=0}^{\infty} \left[M_s \ln\left(1 + \frac{N_s - 1}{M_s}\right) + (N_s - 1)\ln\left(\frac{M_s}{N_s - 1} + 1\right)\right]$$

[see Eq.(19)]

$$= k_B \textstyle\sum_{s=0}^{\infty} \left[M_s \zeta h\nu_s + (N_s - 1)\ln\left(\frac{e^{\zeta h\nu_s}}{e^{\zeta h\nu_s} - 1}\right)\right]$$

[$N_s - 1 \cong N_s$]

$$= k_B \textstyle\sum_{s=0}^{\infty} [\zeta M_s h\nu_s - (N_s - 1)\ln(1 - e^{-\zeta h\nu_s})] \cong k_B[\zeta\mathcal{E} - \sum_{s=0}^{\infty} N_s \ln(1 - e^{-\zeta h\nu_s})]. \qquad (21)$$

The expression of the entropy in Eq.(21) is identical with that obtained previously via Bose's original counting method as seen in the 5th line of Eq.(12).

**4. Einstein's statistically-independent distribution of photons in the phase-space**. In the second part of his paper,[8] Einstein points out a crucial difference between (i) Bose's distribution of $M_s$ identical, indistinguishable photons in the frequency interval $[\nu_s, \nu_s + \mathrm{d}\nu_s)$ among the $N_s$ available cells of the phase-space, and (ii) a statistically-independent distribution of the same number of distinguishable photons among the same cells. In the latter case, Einstein starts with a total of $\mathbb{N}$ distinct particles (e.g., photons), which he then distributes among the various frequency intervals in $\mathbb{N}!/(\prod_{s=0}^{\infty} M_s!)$ different ways — with $\sum_{s=0}^{\infty} M_s = \mathbb{N}$. The $M_s$ (distinguishable) photons thus assigned to the $N_s$ available phase-space cells in the frequency range $[\nu_s, \nu_s + \mathrm{d}\nu_s)$ can assume $(N_s)^{M_s}$ different configurations, as depicted in Fig.2. The number of distinct configurations of the $\mathbb{N}$ photons distributed across the entire range of frequencies $\nu_1, \nu_2, \nu_3, \cdots$ is thus given by

$$w = \mathbb{N}! \textstyle\prod_{s=0}^{\infty} [(N_s)^{M_s}/M_s!]. \qquad (22)$$

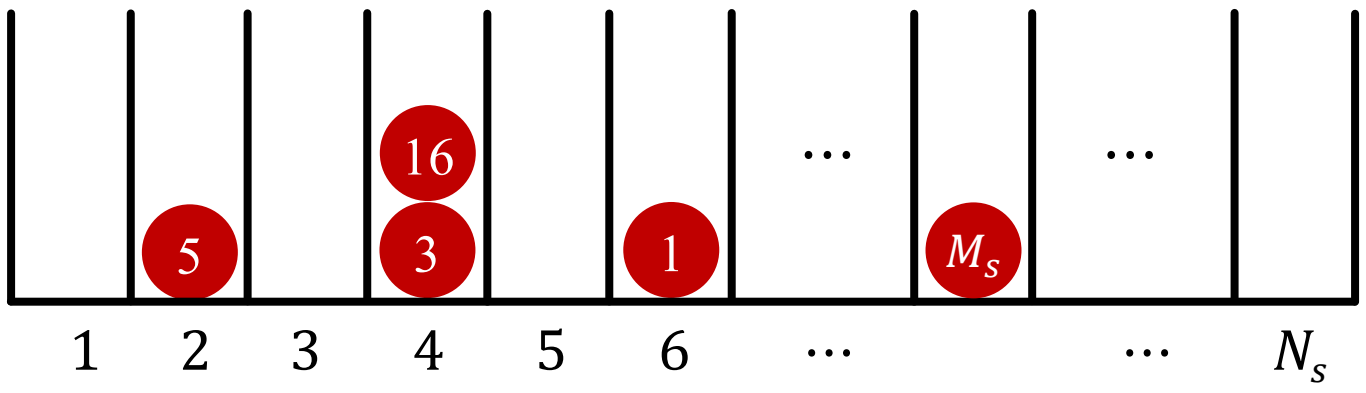

**Fig.2**. A statistically-independent distribution of $M_s$ photons in the frequency range $[\nu_s, \nu_s + \mathrm{d}\nu_s)$ among the corresponding $N_s$ cells of the phase space. Individual photons are taken to be distinct, say, labeled with identifying numbers $1, 2, 3, \cdots, M_s$. Dropping each photon randomly and independently of all the others into a cell will result in $(N_s)^{M_s}$ different arrangements.

Einstein treats $\mathbb{N}!$ as an inconsequential constant and proceeds to ignore it in his subsequent analysis of the photon gas.** Invoking the Stirling[3] approximation $n! \cong (n/e)^n$, one now writes the natural logarithm of $w$ (with the $\mathbb{N}!$ coefficient dropped) as follows:

$$\ln(w) \cong \sum_{s=0}^{\infty}[M_s \ln(N_s) - M_s \ln(M_s) + M_s]. \tag{23}$$

The composite function formed by $\ln(w)$ and the overall energy $\mathcal{E}$ of the photon gas is written as

$$\sum_{s=0}^{\infty}[M_s \ln(N_s) - M_s \ln(M_s) + M_s] - \zeta(\sum_{s=0}^{\infty} M_s h\nu_s). \quad \leftarrow \boxed{\zeta \text{ is the Lagrange multiplier}} \tag{24}$$

Equating to zero the derivatives of the above function with respect to each and every independent variable $M_s$, we arrive at

$$\ln(N_s) - \ln(M_s) - \zeta h\nu_s = 0 \qquad \rightarrow \qquad M_s = N_s e^{-\zeta h\nu_s}. \tag{25}$$

Finding the value of the Lagrange multiplier $\zeta$ now requires enforcing the energy constraint; that is,

$$\mathcal{E} = \sum_{s=0}^{\infty} M_s h\nu_s = \sum_{s=0}^{\infty} N_s e^{-\zeta h\nu_s} h\nu_s = (8\pi hV/c^3)\sum_{s=0}^{\infty} \nu_s^3 e^{-\zeta h\nu_s} \mathrm{d}\nu_s$$
$$= (8\pi hV/c^3)\int_0^{\infty} \nu_s^3 e^{-\zeta h\nu_s}\mathrm{d}\nu_s = \frac{6V}{\pi^2 c^3 \hbar^3 \zeta^4}. \quad \leftarrow \boxed{\text{G\&R } \mathbf{3.351}\text{-3; } \hbar = h/2\pi} \tag{26}$$

The entropy of the photon gas may be evaluated from Eq.(23) with the aid of Eq.(25), as follows:

$$S = k_B \ln(w) \cong k_B \sum_{s=0}^{\infty} M_s[1 + \ln(N_s/M_s)] = k_B \sum_{s=0}^{\infty} N_s e^{-\zeta h\nu_s}(1 + \zeta h\nu_s)$$
$$= (8\pi k_B V/c^3)\int_0^{\infty} \nu_s^2 (1 + \zeta h\nu_s) e^{-\zeta h\nu_s}\mathrm{d}\nu_s = (8\pi k_B V/c^3)\frac{8}{(\zeta h)^3} = \frac{8k_B V}{\pi^2 (c\hbar\zeta)^3}. \tag{27}$$

Finally, the absolute temperature $T$ of the photon gas is derived from the thermodynamic identity involving the energy $\mathcal{E}$ and the entropy $S$ at a fixed volume $V$, namely,

$$T = \frac{\partial\mathcal{E}/\partial\zeta}{\partial S/\partial\zeta} = \frac{(-4)6V}{\pi^2(c\hbar)^3\zeta^5} \div \frac{(-3)8k_B V}{\pi^2(c\hbar)^3\zeta^4} = \frac{1}{k_B\zeta} \qquad \rightarrow \qquad \zeta = 1/(k_B T). \tag{28}$$

Observe that the spectral energy-density of the blackbody radiation as appearing in Eq.(26) is no longer the Planck distribution; rather, it is the Wien displacement law, $(8\pi hV\nu_s^3/c^3)e^{-h\nu_s/k_B T}$, which was a precursor of the Planck distribution. (The Planck constant $h$, of course, did not appear in Wien's original expression of his law; a different symbol was used to represent this constant.)

Both the energy $\mathcal{E}$ of Eq.(26) and the entropy $S$ of Eq.(27) approach zero in the limit of $T \rightarrow 0$. Also, both $\mathcal{E}$ and $S$ are proportional to the volume $V$, which indicates that, if two boxes of volumes $V_1$ and $V_2$ at the same temperature $T$ are conjoined, the energy and the entropy of the combined photon gas will be the sum of the energies and the entropies of the individual boxes, respectively. Let us not forget, however, that in the case of the entropy $S$, the constant term $\ln(\mathbb{N}!) \cong \mathbb{N}\ln(\mathbb{N}) - \mathbb{N}$ has been dropped at the outset from the expression of $\ln(w)$. Had this term been retained, the photon gas entropies in the two boxes would not have been additive (i.e., the Gibbs paradox).[6]

---

** The constraint $\sum_{s=0}^{\infty} M_s = \mathbb{N}$ on the total number of particles is imposed by Einstein in his parallel treatment of an ideal monoatomic gas, but ignored in the case of the photon gas.[8] An alternative interpretation of his "statistically-independent" photon distribution is to ignore $\mathbb{N}$ altogether, treat all $M_s$ photons in the frequency interval $[\nu_s, \nu_s + \mathrm{d}\nu_s)$ as indistinguishable, then take their corresponding number of configurations as $(N_s)^{M_s}/M_s!$. The approximate nature of this counting method is readily appreciated by examining a few cases where $M_s$ and $N_s$ are small integers. For instance, for $(N_s, M_s) = (4, 2)$, the exact number of distinct distributions is 10, whereas $(N_s)^{M_s}/M_s! = 8$, an underestimation by 20%. For $(N_s, M_s) = (10, 2)$, the exact number is 55 while $10^2/2! = 50$, an underestimate of only about 10%. Similarly, for $(N_s, M_s) = (30, 4)$, the exact number is 40,920, whereas $30^4/4! = 33{,}750$ constitutes a 17.5% under-estimate; however, for $(N_s, M_s) = (60, 4)$, the exact number is 595,665, for which $60^4/4! = 540{,}000$ is only a 9.5% underestimate. The approximation becomes far more accurate when $N_s \gg M_s$.

**5. Einstein's quantum theory of the monoatomic ideal gas**. Einstein's treatment of a monoatomic gas consisting of $\mathbb{N}$ identical atoms/molecules in thermal equilibrium with the walls of their closed container (volume $= V$) follows Bose's derivation[1] of the Planck law for a photon gas under similar circumstances.[7,8] The first major difference between the photon gas and the gas consisting of material particles is that the former does not have a specific constraint on the overall number of photons. The second major difference is that each photon is taken to have frequency $\nu$, energy $h\nu$, linear momentum $h\nu/c$ (along the direction of propagation), and one of two distinct states of polarization (e.g., right or left circular), whereas Einstein's identical particles have mass $m$, linear momentum $\boldsymbol{p}$, and (non-relativistic) kinetic energy $\epsilon = p^2/2m$, with no polarization (i.e., angular momentum); thus, $\mathrm{d}\epsilon = p\mathrm{d}p/m$ and $p^2\mathrm{d}p = mp\mathrm{d}\epsilon = m(2m\epsilon)^{½}\mathrm{d}\epsilon$. The infinitesimal volume of the position-momentum phase space that is available to monoatomic ideal gas particles whose energy lies in the narrow interval $[\epsilon_s, \epsilon_s + \mathrm{d}\epsilon_s)$ is thus given by

$$\int \mathrm{d}x\mathrm{d}y\mathrm{d}z\mathrm{d}p_x\mathrm{d}p_y\mathrm{d}p_z = V \times \overbrace{4\pi p^2\mathrm{d}p}^{\text{spherical shell volume}} = 2\pi V(2m)^{3/2}\sqrt{\epsilon_s}\mathrm{d}\epsilon_s. \tag{29}$$

Taking the discretized volume of each elementary cell within the six-dimensional phase space to be $h^3$, the total number of such cells available for occupation by particles whose energy lies in the interval $[\epsilon_s, \epsilon_s + \mathrm{d}\epsilon_s)$ is seen to be

$$N_s = (2\pi V/h^3)(2m)^{3/2}\sqrt{\epsilon_s}\mathrm{d}\epsilon_s. \tag{30}$$

Following Bose's lead, Einstein takes the energy $\epsilon_s$ of individual gas particles to be anywhere in the $[0, \infty)$ interval, proceeding to assign the values $0, 1, 2, 3, \cdots$ to the index $s$. He also assumes that, among the $N_s$ elementary cells associated with the energy interval $[\epsilon_s, \epsilon_s + \mathrm{d}\epsilon_s)$, the number of cells that contain $r$ particles is $n_{r,s}$, with $r$ being $0, 1, 2, 3, \cdots$. Denoting the overall number of particles of the monoatomic ideal gas by $\mathbb{N}$, the total energy content of the box by $\mathcal{E}$, and the number of distinct arrangements of the particles within elementary cells by $w$, the aforementioned assumptions lead to the following identities:

$$N_s = \textstyle\sum_{r=0}^{\infty} n_{r,s}. \tag{31}$$

$$\mathbb{N} = \textstyle\sum_{s=0}^{\infty}\sum_{r=0}^{\infty} r n_{r,s}. \tag{32}$$

$$\mathcal{E} = \textstyle\sum_{s=0}^{\infty}\sum_{r=0}^{\infty} r n_{r,s}\epsilon_s. \tag{33}$$

$$w = \textstyle\prod_{s=0}^{\infty}[N_s!/\prod_{r=0}^{\infty}(n_{r,s}!)]. \tag{34}$$

The goal at this point is to maximize $w$ of Eq.(34) subject to the constraints imposed on $n_{r,s}$ by Eqs.(31)-(33). Invoking the Stirling approximation,[3] $n! \cong (n/e)^n$, the natural logarithm of $w$ may be reduced to a couple of infinite sums over $r$ and $s$, as follows: **Note**: Eq.(31) has been invoked here.

$$\ln(w) = \textstyle\sum_{s=0}^{\infty}\ln(N_s!) - \sum_{s=0}^{\infty}\sum_{r=0}^{\infty}\ln(n_{r,s}!) \cong \sum_{s=0}^{\infty} N_s\ln(N_s) - \sum_{s=0}^{\infty}\sum_{r=0}^{\infty} n_{r,s}\ln(n_{r,s}). \tag{35}$$

Considering that $N_s$ appearing on the right-hand side of Eq.(35) is fixed by Eq.(30) and is, therefore, independent of the variables $n_{r,s}$, one needs only to minimize the second term in the above expression of $\ln(w)$ subject to the constraints of Eqs.(31)-(33). Forming the composite function

$$\sum_{s=0}^{\infty}\sum_{r=0}^{\infty} n_{r,s}\ln(n_{r,s}) + \lambda_s\left(\sum_{r=0}^{\infty} n_{r,s}\right) + \eta\left(\sum_{s=0}^{\infty}\sum_{r=0}^{\infty} r n_{r,s}\right) + \zeta\left(\sum_{s=0}^{\infty}\sum_{r=0}^{\infty} r n_{r,s}\epsilon_s\right), \tag{36}$$

Lagrange multipliers

we proceed to set its derivatives with respect to each and every $n_{r,s}$ equal to zero.[3] We thus arrive at

$$[\ln(n_{r,s}) + 1] + \lambda_s + \eta r + \zeta r\epsilon_s = 0 \qquad \rightarrow \qquad n_{r,s} = e^{-(1+\lambda_s)}e^{-(\eta+\zeta\epsilon_s)r}. \tag{37}$$

The values of the Lagrange multipliers $\lambda_s$ may now be obtained by enforcing the constraints of Eq.(31); that is,

assuming $\eta$ and $\zeta$ are positive

$$N_s = \sum_{r=0}^{\infty} n_{r,s} = e^{-(1+\lambda_s)}\sum_{r=0}^{\infty} e^{-(\eta+\zeta\epsilon_s)r} = e^{-(1+\lambda_s)}/[1 - e^{-(\eta+\zeta\epsilon_s)}]$$

$$\rightarrow \quad n_{r,s} = N_s[1 - e^{-(\eta+\zeta\epsilon_s)}]e^{-(\eta+\zeta\epsilon_s)r}. \tag{38}$$

The constraints imposed by Eq.(32) on the total number $\mathbb{N}$ of the particles, and by Eq.(33) on the overall kinetic energy $\mathcal{E}$ of the particles enclosed within a box of volume $V$, now yield the following pair of coupled equations for the Lagrange multipliers $\eta$ and $\zeta$:

$$\mathbb{N} = \sum_{s=0}^{\infty} N_s[1 - e^{-(\eta+\zeta\epsilon_s)}]\sum_{r=0}^{\infty} r e^{-(\eta+\zeta\epsilon_s)r} = \sum_{s=0}^{\infty} N_s e^{-(\eta+\zeta\epsilon_s)}/[1 - e^{-(\eta+\zeta\epsilon_s)}]$$

$x/(1-x) = \sum_{n=1}^{\infty} x^n$ ; $\sum_{r=0}^{\infty} r x^r = x/(1-x)^2$

$$= \sum_{s=0}^{\infty} N_s \sum_{n=1}^{\infty} e^{-n(\eta+\zeta\epsilon_s)} = (2\pi V/h^3)(2m)^{3/2}\sum_{n=1}^{\infty} e^{-n\eta}\int_0^{\infty}\sqrt{\epsilon_s}e^{-n\zeta\epsilon_s}\mathrm{d}\epsilon_s$$

$$= (2\pi V/h^3)(2m/\zeta)^{3/2}\Gamma(3/2)\sum_{n=1}^{\infty}(e^{-n\eta}/n^{3/2}). \quad \leftarrow \text{G\&R } \mathbf{3.381}\text{-4} \tag{39}$$

$$\mathcal{E} = \sum_{s=0}^{\infty} N_s\epsilon_s[1 - e^{-(\eta+\zeta\epsilon_s)}]\sum_{r=0}^{\infty} r e^{-(\eta+\zeta\epsilon_s)r} = \sum_{s=0}^{\infty} N_s\epsilon_s e^{-(\eta+\zeta\epsilon_s)}/[1 - e^{-(\eta+\zeta\epsilon_s)}]$$

$$= (2\pi V/h^3)(2m)^{3/2}\sum_{n=1}^{\infty} e^{-n\eta}\int_0^{\infty}\epsilon_s^{3/2}e^{-n\zeta\epsilon_s}\mathrm{d}\epsilon_s$$

$$= (2\pi V/h^3)(2m)^{3/2}\Gamma(5/2)\zeta^{-5/2}\sum_{n=1}^{\infty}(e^{-n\eta}/n^{5/2}). \leftarrow \text{G\&R } \mathbf{3.381}\text{-4} \tag{40}$$

Equations (39) and (40) can be solved numerically for the values of the Lagrange multipliers $\eta$ and $\zeta$ in terms of the number $\mathbb{N}$ of the particles and the energy $\mathcal{E}$ of the confined monoatomic gas. It is possible, however, to relate $\zeta$ to the absolute temperature $T$ of the gas without explicitly solving these equations. To this end, we use Eq.(35) to express the entropy $S$ of the gas in terms of $\eta$ and $\zeta$, as follows:

$$S = k_B\ln(w) \cong k_B\left\{\sum_{s=0}^{\infty} N_s\ln(N_s) - \sum_{s=0}^{\infty}\sum_{r=0}^{\infty} n_{r,s}\ln(n_{r,s})\right\}$$

$$= k_B\left\{\sum_{s=0}^{\infty} N_s\ln(N_s) + \sum_{s=0}^{\infty}\sum_{r=0}^{\infty} n_{r,s}(1 + \lambda_s + \eta r + \zeta r\epsilon_s)\right\} \quad \leftarrow \text{see Eq.(37)}$$

$$= k_B\left\{\sum_{s=0}^{\infty} N_s\ln(N_s) + \sum_{s=0}^{\infty}(1 + \lambda_s)N_s + \sum_{s=0}^{\infty}(\eta + \zeta\epsilon_s)\sum_{r=0}^{\infty} r n_{r,s}\right\} \quad \leftarrow \text{see Eq.(38)}$$

$\ln(1-x) = -\sum_{n=1}^{\infty}(x^n/n)$ ; $\sum_{r=0}^{\infty} r x^r = x/(1-x)^2$

$$= k_B\left\{-\sum_{s=0}^{\infty} N_s\ln[1 - e^{-(\eta+\zeta\epsilon_s)}] + \sum_{s=0}^{\infty}(\eta + \zeta\epsilon_s)N_s e^{-(\eta+\zeta\epsilon_s)}/[1 - e^{-(\eta+\zeta\epsilon_s)}]\right\}$$

$x/(1-x) = \sum_{n=1}^{\infty} x^n$

$$= k_B(2\pi V/h^3)(2m)^{3/2}\left[\sum_{n=1}^{\infty} n^{-1}e^{-n\eta}\int_0^{\infty}\sqrt{\epsilon_s}e^{-n\zeta\epsilon_s}\mathrm{d}\epsilon_s + \sum_{n=1}^{\infty} e^{-n\eta}\int_0^{\infty}(\eta + \zeta\epsilon_s)\sqrt{\epsilon_s}e^{-n\zeta\epsilon_s}\mathrm{d}\epsilon_s\right]$$

G&R **3.381**-4 ; G&R **3.381**-4

$$= k_B(2\pi V/h^3)(2m/\zeta)^{3/2}\{[\Gamma(3/2)+\Gamma(5/2)]\textstyle\sum_{n=1}^{\infty}(e^{-n\eta}/n^{5/2})+\eta\Gamma(3/2)\sum_{n=1}^{\infty}(e^{-n\eta}/n^{3/2})\}$$

$$= k_B(2\pi V/h^3)(2m/\zeta)^{3/2}\big[(5\sqrt{\pi}/4)\textstyle\sum_{n=1}^{\infty}(e^{-n\eta}/n^{5/2})+½\sqrt{\pi}\eta\sum_{n=1}^{\infty}(e^{-n\eta}/n^{3/2})\big]. \quad (41)$$

Having found expressions for $\mathbb{N}$, $\mathcal{E}$, and $S$ as functions of $V$, $\eta$, and $\zeta$, we proceed to simplify the notation by introducing three new functions, as follows:

$$g_1(\eta)=\textstyle\sum_{n=1}^{\infty}(e^{-n\eta}/n^{1/2}), \quad g_3(\eta)=\sum_{n=1}^{\infty}(e^{-n\eta}/n^{3/2}), \quad g_5(\eta)=\sum_{n=1}^{\infty}(e^{-n\eta}/n^{5/2}). \quad (42)$$

Fixing the volume $V$, allowing for infinitesimal variations $\delta\eta, \delta\zeta$ in the $\eta, \zeta$ parameters, and ensuring that the number $\mathbb{N}$ of the particles remains unchanged, we will have

$$\delta\mathbb{N}=0 \;\rightarrow\; -(3/2)\zeta^{-5/2}g_3(\eta)\delta\zeta-\zeta^{-3/2}g_1(\eta)\delta\eta=0 \;\rightarrow\; \zeta g_1(\eta)\delta\eta=-(3/2)g_3(\eta)\delta\zeta. \quad (43)$$

$$\delta\mathcal{E}=¾\sqrt{\pi}(2\pi V/h^3)(2m)^{3/2}\big[-(5/2)\zeta^{-7/2}g_5(\eta)\delta\zeta-\zeta^{-5/2}g_3(\eta)\delta\eta\big]$$

see Eq.(43) → $$=⅜\sqrt{\pi}(2\pi V/h^3)(2m)^{3/2}[-5g_5(\eta)+3g_3^2(\eta)/g_1(\eta)]\zeta^{-7/2}\delta\zeta. \quad (44)$$

$$\delta S=-(3/2)k_B(2\pi V/h^3)(2m)^{3/2}\zeta^{-5/2}\big[(5\sqrt{\pi}/4)g_5(\eta)+½\sqrt{\pi}\eta g_3(\eta)\big]\delta\zeta$$
$$+k_B(2\pi V/h^3)(2m/\zeta)^{3/2}\big[-(3\sqrt{\pi}/4)g_3(\eta)-½\sqrt{\pi}\eta g_1(\eta)\big]\delta\eta$$ ← see Eq.(43)
$$=-(3/2)k_B(2\pi V/h^3)(2m)^{3/2}\big[(5\sqrt{\pi}/4)g_5(\eta)+½\sqrt{\pi}\eta g_3(\eta)\big]\zeta^{-5/2}\delta\zeta$$
$$+(3/2)k_B(2\pi V/h^3)(2m)^{3/2}\big[(3\sqrt{\pi}/4)\,g_3^2(\eta)/g_1(\eta)+½\sqrt{\pi}\eta g_3(\eta)\big]\zeta^{-5/2}\delta\zeta$$
$$=⅜\sqrt{\pi}k_B(2\pi V/h^3)(2m)^{3/2}[-5g_5(\eta)+3g_3^2(\eta)/g_1(\eta)]\zeta^{-5/2}\delta\zeta. \quad (45)$$

Invoking the thermodynamic identity $\delta S/\delta\mathcal{E}=1/T$, we find $\zeta=1/(k_BT)$. Consequently,

$$\mathbb{N}=(V/h^3)(2\pi m)^{3/2}(k_BT)^{3/2}g_3(\eta). \quad (46)$$

$$\mathcal{E}=(3V/2h^3)(2\pi m)^{3/2}(k_BT)^{5/2}g_5(\eta). \quad (47)$$

$$S=(5\mathcal{E}/3T)+\eta k_B\mathbb{N}. \quad (48)$$

Helmholtz free energy → $$F=\mathcal{E}-Ts=-⅔\mathcal{E}-\eta k_BT\mathbb{N}. \quad (49)$$

**5.1. Pressure**. We evaluate the pressure $\wp$ of the gas via the thermodynamic identity $\wp=-\partial F/\partial V$ using the infinitesimal variations $\delta V$ and $\delta\eta$ in the gas volume $V$ and the parameter $\eta$, while fixing the temperature $T$ and the number $\mathbb{N}$ of the particles. We will have

$$\delta\mathbb{N}=0 \quad\rightarrow\quad g_3(\eta)\delta V=Vg_1(\eta)\delta\eta. \quad (50)$$

$$\delta\mathcal{E}=(3/2)(2\pi m/h^2)^{3/2}(k_BT)^{5/2}[\delta Vg_5(\eta)-Vg_3(\eta)\delta\eta]$$
$$=(3/2)(2\pi m/h^2)^{3/2}(k_BT)^{5/2}[g_5(\eta)-g_3^2(\eta)/g_1(\eta)]\delta V. \quad (51)$$

Therefore,

$$\wp=-\partial F/\partial V=(2\pi m/h^2)^{3/2}(k_BT)^{5/2}[g_5(\eta)-g_3^2(\eta)/g_1(\eta)]$$ ← see Eq.(49)
$$+[k_BTg_3(\eta)/Vg_1(\eta)]\big[(2\pi mk_BT/h^2)^{3/2}Vg_3(\eta)\big]$$ ← see Eqs.(46), (49), (50)
$$\rightarrow\quad \wp=(2\pi m/h^2)^{3/2}(k_BT)^{5/2}g_5(\eta)=⅔(\mathcal{E}/V). \quad (52)$$

**5.2. Specific heat at constant volume**. One can determine $C_V$, the specific heat at a constant volume of the monoatomic ideal gas, by fixing $V$ and $\eta$, then computing $\partial\mathcal{E}/\partial T$, as follows:

$$\delta\mathbb{N} = 0 \;\rightarrow\; (3/2)T^{1/2}g_3(\eta)\delta T + T^{3/2}g_3'(\eta)\delta\eta = 0 \;\rightarrow\; \delta\eta = -[3g_3(\eta)/2g_3'(\eta)]\,\delta T/T. \quad (53)$$

$$\begin{aligned}\delta\mathcal{E} &= (3V/2h^3)(2\pi m)^{3/2}k_B^{5/2}\big[(5/2)T^{3/2}g_5(\eta)\delta T + T^{5/2}g_5'(\eta)\delta\eta\big]\\ &= (3k_BV/2h^3)(2\pi m)^{3/2}(k_BT)^{3/2}[(5/2)g_5(\eta) - (3/2)g_5'(\eta)g_3(\eta)/g_3'(\eta)]\delta T\\ &= (3k_B\mathbb{N}/2)[(5/2)g_5(\eta)/g_3(\eta) - (3/2)g_5'(\eta)/g_3'(\eta)]\delta T. \end{aligned} \quad (54)$$

Taking $\mathbb{N}$ to be Avogadro's number $N_A = 6.02214076 \times 10^{23}$ mol$^{-1}$, and using the universal gas constant $R = k_BN_A \cong 8.314$ Joule/(mol $\cdot K$), the specific heat per mole of the gas amounts to

$$C_V = \frac{\partial\mathcal{E}}{\partial T} = \frac{3R}{2}\left\{\frac{g_5(\eta)}{g_3(\eta)} - \frac{3}{2}\left[\frac{g_5'(\eta)}{g_3'(\eta)} - \frac{g_5(\eta)}{g_3(\eta)}\right]\right\} = \frac{3R}{2}\left\{\frac{g_5(\eta)}{g_3(\eta)} - \frac{3}{2}\frac{g_3(\eta)}{g_3'(\eta)}\left[\frac{g_5(\eta)}{g_3(\eta)}\right]'\right\}. \quad (55)$$

Needless to say, one could as well substitute $g_3(\eta)/g_1(\eta)$ for $g_5'(\eta)/g_3'(\eta)$ in Eqs.(54) and (55).††

**5.3. Specific heat at constant pressure**. The specific heat at constant pressure, $C_{\wp}$, is evaluated by fixing the number of particles $\mathbb{N}$, and also fixing the pressure $\wp$ — i.e., fixing the energy-density $\mathcal{E}/V$, in accordance with Eq.(52) — then computing the infinitesimal amount of supplied energy, $\delta\mathcal{E} + \wp\delta V$, needed to raise the temperature by $\delta T$ while the volume simultaneously increases by $\delta V$. We will have

$$\delta\wp = 0 \;\rightarrow\; (5/2)T^{3/2}g_5(\eta)\delta T - T^{5/2}g_3(\eta)\delta\eta = 0 \;\rightarrow\; \delta\eta = (5/2)[g_5(\eta)/g_3(\eta)]\,\delta T/T. \quad (56)$$

$$\begin{aligned}\delta\mathbb{N} = 0 \;&\rightarrow\; T^{3/2}g_3(\eta)\delta V + (3/2)VT^{1/2}g_3(\eta)\delta T - VT^{3/2}g_1(\eta)\delta\eta = 0\\ &\rightarrow\; \delta V/V = \{(5/2)[g_1(\eta)g_5(\eta)/g_3^2(\eta)] - (3/2)\}\,\delta T/T. \leftarrow \boxed{\text{see Eq.(56)}}\end{aligned} \quad (57)$$

$$\begin{aligned}\delta\mathcal{E} &= (3/2h^3)(2\pi m)^{3/2}k_B^{5/2}\big[T^{5/2}g_5(\eta)\delta V + (5/2)VT^{3/2}g_5(\eta)\delta T - VT^{5/2}g_3(\eta)\delta\eta\big]\\ &= (3k_BV/2h^3)(2\pi m)^{3/2}(k_BT)^{3/2}[(5/2)\,g_1(\eta)g_5^2(\eta)/g_3^2(\eta) - (3/2)g_5(\eta)]\delta T.\end{aligned} \quad (58)$$

$$\wp\delta V = k_BV(2\pi m/h^2)^{3/2}(k_BT)^{3/2}[(5/2)\,g_1(\eta)g_5^2(\eta)/g_3^2(\eta) - (3/2)g_5(\eta)]\delta T. \quad (59)$$

Upon combining Eqs.(58) and (59), substituting for $\mathbb{N}$ from Eq.(46), taking $\mathbb{N}$ to be Avogadro's number $N_A$, and recalling that $k_BN_A$ is the universal gas constant $R$, we arrive at

$$C_{\wp} = (\delta\mathcal{E} + \wp\delta V)/\delta T = \frac{5R}{2}\left[\frac{5g_1(\eta)g_5^2(\eta)}{2g_3^3(\eta)} - \frac{3g_5(\eta)}{2g_3(\eta)}\right], \qquad \text{Joule/(mol}\cdot K). \quad (60)$$

**5.4. Discussion**. Note that the function $g_1(\eta)$ defined in Eq.(42) is divergent for $\eta \leq 0$, and that $g_3(\eta)$ and $g_5(\eta)$ also diverge if $\eta < 0$. All three functions are well-defined for $\eta > 0$ and decline with an increasing $\eta$. While $g_1(0) = \infty$, the values of the other two functions at $\eta = 0$ are $g_3(0) \cong 2.6124$ and $g_5(0) \cong 1.3415$. According to Eq.(46), at any given temperature $T$, the number of particles per unit volume, $\mathbb{N}/V$, has an upper limit, as $g_3(\eta)$ reaches its peak value at $\eta = 0$. Einstein conjectured that the number-density of the particles cannot exceed this critical (or saturation) value, and that any attempt at increasing the particles' number-density at a given temperature will result in the additional particles separating from the rest and condensing into the lowest quantum state with zero kinetic energy. This is similar to what happens when isothermally compressing a vapor beyond its volume of saturation. According to Einstein, an analogue of "oversaturated vapor" does not exist for the ideal gas. He also asserted (in a footnote) that the "condensed" part of the substance claims no particular volume, since it contributes nothing to the pressure.[8]

†† Our expression of $C_V$ is nearly the same as that in Einstein's original paper,[8] except for the ratio $g_3(\eta)/g_3'(\eta)$ on the right-hand side of Eq.(55), which shows up without its denominator in Einstein's paper.

At a given volume $V$ and given number $\mathbb{N}$ of particles, as one lowers the temperature $T$, the value of $\eta$ eventually reaches zero, at which point condensation begins, namely, the excess particles lose their kinetic energy and move into the lowest quantum state, forming a condensate. Thus, in the limit of $T \to 0$, the energy $\mathcal{E}$, the pressure $\wp$, and the entropy $S$ must all vanish. Einstein writes: "According to the theory developed here, Nernst's theorem[‡‡] is satisfied for ideal gases. It has to be kept in mind that our formulae cannot be applied to extremely low temperatures, because in their derivation we have assumed that the $p_r^s$ [related to $n_{r,s}$ in our notation] vary only by a relatively infinitesimally small amount when $s$ varies by 1. On the other hand one recognizes immediately that the entropy has to vanish at the absolute zero point of temperature. Namely, then all molecules are assembled in the first cell; for this state there is only a single distribution of the molecules when adopting our way of counting. From this follows immediately the correctness of our assertion."[7]

Einstein notes that, in its saturated state, the gas has $\eta = 0$ and $S = (\mathcal{E} + \wp V)/T$; see Eqs.(48) and (52). The saturated gas is, of course, in thermal equilibrium with the condensate, for which the energy $\mathcal{E}$, the entropy $S$, and the pressure $\wp$ have all vanished. He goes on to express his belief that Bose's statistical ansatz "has to be preferred, even if the preference of this method to others cannot be proven a priori. This result, in its turn, provides support for the notion of a deep inner relationship between radiation and gas, insofar as the same statistical viewpoint which leads to Planck's formula establishes, when applied to ideal gases, the agreement of the gas theory with Nernst's theorem."[8]

Einstein considered it remarkable that the average kinetic energy per particle obtained from Eqs.(46) and (47) is $\mathcal{E}/\mathbb{N} = (3k_BT/2)g_5(\eta)/g_3(\eta)$, and that the pressure of the gas, in accordance with Eq.(52), is $\wp = ⅔(\mathcal{E}/V)$. While the latter property fully agrees with the results of classical thermodynamics, the average kinetic energy of the particles according to the Bose-Einstein statistics is lower than its classical Maxwell-Boltzmann value of $3k_BT/2$ by a factor of $g_5(\eta)/g_3(\eta)$. This reduction factor is about 0.5135 when the gas is close to saturation (i.e., when $\eta \to 0$), but is also non-negligible at low particle densities. For sufficiently large values of $\eta$ where $e^{-\eta} \ll 1$, it is easy to show that $g_3(\eta) \cong g_5(\eta) \cong e^{-\eta}$ and $g_5(\eta)/g_3(\eta) \cong 1 - (e^{-\eta}/4\sqrt{2})$, and that, therefore,[§§]

$$\mathcal{E}/\mathbb{N} \cong (3k_BT/2)\left[1 - 0.1768h^3(\mathbb{N}/V)(2\pi mk_BT)^{-3/2}\right]. \qquad (61)$$

This kind of deviation from the Maxwell-Boltzmann statistics is also seen in Eq.(39), where the number of particles in the infinitesimal energy interval $[\epsilon_s, \epsilon_s + \mathrm{d}\epsilon_s)$ is given by

$$N_s \textstyle\sum_{n=1}^{\infty} e^{-n(\eta+\zeta\epsilon_s)} = (2\pi V/h^3)(2m)^{3/2}\epsilon_s^{½}e^{-\eta}\left[1 + e^{-(\eta+\epsilon_s/k_BT)} + \cdots\right]e^{-\epsilon_s/k_BT}\mathrm{d}\epsilon_s. \qquad (62)$$

Commenting on the appearance of the bracketed term in Eq.(62), Einstein attributes it to the influence of the "quanta" (as related to Bose's specific counting method) on Maxwell's distribution law: "One sees that the slower molecules occur more frequently, as compared to the fast ones, than they would by virtue of Maxwell's law."[7]

---

‡‡ Nernst's theorem states that the entropy vanishes at zero absolute temperature.

§§ In Einstein's first paper, the coefficient 0.1768 appearing in our Eq.(61) is incorrectly written as 0.0318. The mistake is corrected in the second paper,[8] although a misprint has crept into the final equation of the second paper as well, where the coefficient is written as 0.186. The correct value (0.1768) appears a little after Eq.(44) of Einstein's second paper.

**6. Einstein's alternative (but equivalent) counting method for the monoatomic gas**. Einstein's alternative (albeit equivalent) counting method[8] leads to the same expressions for $\mathbb{N}$, $\mathcal{E}$, and $S$, as demonstrated below. In this counting, the total number of distinct configurations is given by

$$w = \prod_{s=0}^{\infty} (M_s + N_s - 1)!/[M_s!\,(N_s - 1)!]. \quad (63)$$

Invoking Stirling's approximation, $n! \cong (n/e)^n$, the natural logarithm of $w$ may be written as

$$\ln(w) = \sum_{s=0}^{\infty}[\ln(M_s + N_s - 1)! - \ln(M_s!) - \ln(N_s - 1)!]$$

$$\cong \sum_{s=0}^{\infty}[(M_s + N_s - 1)\ln(M_s + N_s - 1) - M_s \ln(M_s) - \overbrace{(N_s - 1)\ln(N_s - 1)}^{\text{independent of } M_s}]. \quad (64)$$

The composite function incorporating $\ln(w)$ and the Lagrange multipliers is

$$\sum_{s=0}^{\infty}[(M_s + N_s - 1)\ln(M_s + N_s - 1) - M_s \ln(M_s)] - \eta \sum_{s=0}^{\infty} M_s - \zeta \sum_{s=0}^{\infty}(M_s \epsilon_s). \quad (65)$$

Setting to zero the derivative of the above function with respect to $M_s$ yields

$$\ln(M_s + N_s - 1) - \ln(M_s) - \eta - \zeta\epsilon_s = 0 \quad \rightarrow \quad M_s = (N_s - 1)/(e^{\eta+\zeta\epsilon_s} - 1). \quad (66)$$

The Lagrange multipliers $\eta$ and $\zeta$ are found by enforcing the constraints on $\mathbb{N}$ and $\mathcal{E}$, as follows:

$$\mathbb{N} = \sum_{s=0}^{\infty} M_s \cong \sum_{s=0}^{\infty}(2\pi V/h^3)(2m)^{3/2}\sqrt{\epsilon_s}(e^{\eta+\zeta\epsilon_s} - 1)^{-1}\mathrm{d}\epsilon_s$$

$$= (2\pi V/h^3)(2m)^{3/2}\int_0^{\infty}\sqrt{\epsilon_s}(e^{\eta+\zeta\epsilon_s} - 1)^{-1}\mathrm{d}\epsilon_s$$

$$= (2\pi V/h^3)(2m)^{3/2}\sum_{n=1}^{\infty}\int_0^{\infty}\epsilon_s^{1/2}e^{-n(\eta+\zeta\epsilon_s)}\mathrm{d}\epsilon_s$$

$$= (2\pi V/h^3)(2m/\zeta)^{3/2}\Gamma(3/2)\sum_{n=1}^{\infty}\left(e^{-n\eta}/n^{3/2}\right) = (V/h^3)(2\pi m/\zeta)^{3/2}g_3(\eta). \quad (67)$$

$$\mathcal{E} = \sum_{s=0}^{\infty}(M_s\epsilon_s) \cong \sum_{s=0}^{\infty} N_s(e^{\eta+\zeta\epsilon_s} - 1)^{-1}\epsilon_s$$

$$\cong \sum_{s=0}^{\infty}(2\pi V/h^3)(2m)^{3/2}\epsilon_s^{3/2}\mathrm{d}\epsilon_s\sum_{n=1}^{\infty} e^{-n(\eta+\zeta\epsilon_s)}$$

$$= (2\pi V/h^3)(2m)^{3/2}\sum_{n=1}^{\infty} e^{-n\eta}\int_0^{\infty}\epsilon_s^{3/2}e^{-n\zeta\epsilon_s}\mathrm{d}\epsilon_s$$

$$= (2\pi V/h^3)(2m)^{3/2}\zeta^{-5/2}\Gamma(5/2)\sum_{n=1}^{\infty}\left(e^{-n\eta}/n^{5/2}\right)$$

$$= (3V/2h^3)(2\pi m)^{3/2}\zeta^{-5/2}g_5(\eta). \quad (68)$$

The entropy of the ideal monoatomic gas is found from Eq.(64) with the aid of Eq.(66), as follows:

$$S = k_B\ln(w) \cong k_B\sum_{s=0}^{\infty}[(M_s + N_s - 1)\ln(M_s + N_s - 1) - M_s\ln(M_s) - (N_s - 1)\ln(N_s - 1)]$$

$$= k_B\sum_{s=0}^{\infty}\left[M_s\ln\left(1 + \frac{N_s-1}{M_s}\right) + (N_s - 1)\ln\left(\frac{M_s}{N_s-1} + 1\right)\right]$$

$$= k_B\sum_{s=0}^{\infty}\{(\eta + \zeta\epsilon_s)M_s - (N_s - 1)\ln[1 - e^{-(\eta+\zeta\epsilon_s)}]\}$$

$$= k_B\sum_{s=0}^{\infty}(N_s \underset{\boxed{N_s - 1 \cong N_s}}{\underset{\uparrow}{-}} 1)\left[(\eta + \zeta\epsilon_s)\sum_{n=1}^{\infty} e^{-n(\eta+\zeta\epsilon_s)} + \sum_{n=1}^{\infty} n^{-1}e^{-n(\eta+\zeta\epsilon_s)}\right]$$

$$\cong (2\pi k_B V/h^3)(2m)^{3/2}\left[\sum_{n=1}^{\infty} e^{-n\eta}\int_0^{\infty}(\eta + \zeta\epsilon_s)\epsilon_s^{1/2}e^{-n\zeta\epsilon_s}\mathrm{d}\epsilon_s + \sum_{n=1}^{\infty} n^{-1}e^{-n\eta}\int_0^{\infty}\epsilon_s^{1/2}e^{-n\zeta\epsilon_s}\mathrm{d}\epsilon_s\right]$$

$$= (2\pi k_B V/h^3)(2m/\zeta)^{3/2}\{[\Gamma(5/2) + \Gamma(3/2)]\sum_{n=1}^{\infty}(e^{-n\eta}/n^{5/2}) + \eta\Gamma(3/2)\sum_{n=1}^{\infty}(e^{-n\eta}/n^{3/2})\}$$

$$= (k_B V/h^3)(2\pi m/\zeta)^{3/2}[(5/2)g_5(\eta) + \eta g_3(\eta)]. \quad (69)$$

The above results are identical with those obtained using Bose's original counting method.

**7. Statistically-independent distribution of particles in a monoatomic gas**. As an alternative to Bose's distribution of particles in a box of volume $V$ among the phase-space cells, Einstein suggested a "statistically-independent" way of distributing $\mathbb{N}$ particles among all the available cells of the phase space.[8] His alternative method entails a division of $\mathbb{N}$ *distinguishable* particles into groups of $M_s$ particles ($\mathbb{N} = \sum_{s=0}^{\infty} M_s$), where the kinetic energy of each particle in group $s$ is in the interval $[\epsilon_s, \epsilon_s + \mathrm{d}\epsilon_s)$. Each such group of $M_s$ particles is subsequently distributed in a statistically-independent way among the $N_s$ available cells in the corresponding energy range. The number of distinct configurations (i.e., particle arrangements within the phase-space cells) thus obatined is

$$w = \mathbb{N}! \prod_{s=0}^{\infty}[(N_s)^{M_s}/M_s!]. \quad (70)$$

Ignoring the (inconsequential) constant factor $\mathbb{N}!$ and invoking the Stirling approximation, $n! \cong (n/e)^n$, one may write the natural logarithm of $w$ (with its coefficient $\mathbb{N}!$ dropped) as follows:

$$\ln(w) \cong \sum_{s=0}^{\infty}[M_s \ln(N_s) - M_s \ln(M_s) + M_s]. \quad (71)$$

The composite function pertaining to the method of Lagrange multipliers is now given by

$$\sum_{s=0}^{\infty}[M_s \ln(N_s) - M_s \ln(M_s) + M_s] - \eta \sum_{s=0}^{\infty} M_s - \zeta \sum_{s=0}^{\infty}(M_s \epsilon_s). \quad (72)$$

Setting to zero the derivatives of the above function with respect to each and every $M_s$, we find

$$\ln(N_s) - \ln(M_s) - \eta - \zeta\epsilon_s = 0 \quad \rightarrow \quad M_s = N_s e^{-(\eta+\zeta\epsilon_s)}. \quad (73)$$

The Lagrange multipliers $\eta$ and $\zeta$ are found by enforcing the constraints on $\mathbb{N}$ and $\mathcal{E}$, as follows:

$$\mathbb{N} = \sum_{s=0}^{\infty} M_s = \sum_{s=0}^{\infty} N_s e^{-(\eta+\zeta\epsilon_s)} = (2\pi V/h^3)(2m)^{3/2} e^{-\eta} \int_0^{\infty} \epsilon_s^{1/2} e^{-\zeta\epsilon_s} \mathrm{d}\epsilon_s$$ ← see Eq.(30)

$$= (2\pi V/h^3)(2m/\zeta)^{3/2}\Gamma(3/2)e^{-\eta} = (V/h^3)(2\pi m/\zeta)^{3/2} e^{-\eta}.$$ ← G&R **3.381**-4 (74)

$$\mathcal{E} = \sum_{s=0}^{\infty}(M_s\epsilon_s) = \sum_{s=0}^{\infty} N_s e^{-(\eta+\zeta\epsilon_s)}\epsilon_s = (2\pi V/h^3)(2m)^{3/2} e^{-\eta} \int_0^{\infty} \epsilon_s^{3/2} e^{-\zeta\epsilon_s} \mathrm{d}\epsilon_s$$

$$= (2\pi V/h^3)(2m)^{3/2}\zeta^{-5/2}\Gamma(5/2)e^{-\eta} = (3V/2h^3)(2\pi m)^{3/2}\zeta^{-5/2}e^{-\eta}.$$ ← G&R **3.381**-4 (75)

The entropy of the gas is now found from Eq.(71) with the aid of Eq.(73); that is,

$$S = k_B \ln(w) \cong k_B \sum_{s=0}^{\infty}[M_s \ln(N_s) - M_s \ln(M_s) + M_s] = k_B \sum_{s=0}^{\infty} M_s[1 + \ln(N_s/M_s)]$$

$$= (2\pi k_B V/h^3)(2m)^{3/2} \int_0^{\infty} \epsilon_s^{1/2}[1 + (\eta + \zeta\epsilon_s)]e^{-(\eta+\zeta\epsilon_s)} \mathrm{d}\epsilon_s$$ ← G&R **3.381**-4

$$= (2\pi k_B V/h^3)(2m/\zeta)^{3/2}[(1+\eta)\Gamma(3/2) + \Gamma(5/2)]e^{-\eta}$$

$$= (k_B V/h^3)(2\pi m/\zeta)^{3/2}[\eta + (5/2)]e^{-\eta}. \quad (76)$$

The Lagrange multipliers $\eta$ and $\zeta$ can be obtained from Eqs.(74) and (75) in terms of $\mathbb{N}$ and $\mathcal{E}$. Even without such calculations, however, it is easy to see that $\mathcal{E} = 3\mathbb{N}/(2\zeta)$ and that, therefore, $\delta\mathcal{E} = -3\mathbb{N}\delta\zeta/(2\zeta^2)$, when the number $\mathbb{N}$ of the particles is kept constant. In addition, fixing $\mathbb{N}$ and the volume $V$ yields, in accordance with Eq.(74),

$$\delta\mathbb{N} = 0 \quad \rightarrow \quad -(3/2)e^{-\eta}\zeta^{-5/2}\delta\zeta - \zeta^{-3/2}e^{-\eta}\delta\eta = 0 \quad \rightarrow \quad \delta\eta = -(3/2)\,\delta\zeta/\zeta. \quad (77)$$

We are now in a position to compute the infinitesimal change $\delta S$ in the entropy when $V$ and $\mathbb{N}$ are fixed but the overall energy $\mathcal{E}$ is allowed to vary. From Eq.(76), we find

$$\delta S = (k_B V/h^3)(2\pi m/\zeta)^{3/2} e^{-\eta}\{-(3/2)(\delta\zeta/\zeta)[\eta + (5/2)] + \delta\eta - [\eta + (5/2)]\delta\eta\}$$

$$= (k_B V/h^3)(2\pi m/\zeta)^{3/2} e^{-\eta}(-3/2)\,\delta\zeta/\zeta. \quad (78)$$

The thermodynamic identity $T = \delta\mathcal{E}/\delta S$ now yields $\zeta = 1/(k_B T)$. All in all, we have arrived at

$$\mathbb{N}/V = (2\pi m k_B T/h^2)^{3/2} e^{-\eta} \quad \rightarrow \quad \eta = \ln[(2\pi m k_B T/h^2)^{3/2}(V/\mathbb{N})]; \quad (79)$$

$$M_s/\mathbb{N} = 2\epsilon_s^{½} e^{-\epsilon_s/k_B T} \mathrm{d}\epsilon_s/[\sqrt{\pi}(k_B T)^{3/2}]; \quad (80)$$ ← see Eqs.(30) and (73)

$$\mathcal{E}/\mathbb{N} = 3k_B T/2; \quad (81)$$

$$S/\mathbb{N} = k_B[\eta + (5/2)]. \quad (82)$$ ← Sackur-Tetrode equation[6]

These are the well-known equations of classical thermodynamics.[6] In particular, Eq.(80) is the Maxwell-Boltzmann distribution of particles in terms of their energy $\epsilon_s$, with the gas being in thermal equilibrium at temperature $T$. Note that the right-hand side of Eq.(80) integrates to 1, as it should. Given that the Lagrange multiplier $\eta$ in the present case is not constrained, Eq.(79) does *not* impose an upper bound on the number-density of the particles at the temperature $T$.

If a box of volume $V_1$ containing $\mathbb{N}_1$ particles of mass $m$ at temperature $T$ is conjoined with a second box of volume $V_2$ containing $\mathbb{N}_2$ particles of the same mass $m$ and at the same temperature $T$, then the energy of the gas mixture will be $\mathcal{E} = \mathcal{E}_1 + \mathcal{E}_2 = 3(\mathbb{N}_1 + \mathbb{N}_2)k_B T/2$. Similarly, the entropies $S_1$ and $S_2$ will be additive for the gas mixture provided that $\eta_1 = \eta_2$ (i.e., $\mathbb{N}_1/V_1 = \mathbb{N}_2/V_2$). Note, however, that for the entropies to be additive it is imperative for the coefficient $\mathbb{N}!$ of $w$ to be dropped from Eq.(70), lest the Gibbs paradox raises its head again.[6]

The pressure of the ideal monoatomic gas is given by $\wp = -\partial F/\partial V$, which requires that the Helmholtz free energy $F = \mathcal{E} - TS$ be differentiated with respect to $V$ while keeping $\mathbb{N}$ and $T$ constant. From $\mathcal{E} = 3k_B T\mathbb{N}/2$ and $S = k_B\mathbb{N}[\eta + (5/2)]$, we find $\delta\mathcal{E} = 0$ and $\delta S = k_B\mathbb{N}\delta\eta$. Also, setting $\delta\mathbb{N} = 0$ in Eq.(79) leads to $\delta\eta = \delta V/V$. Consequently,

$$\wp = -\partial F/\partial V = T\delta S/\delta V = k_B T\mathbb{N}/V = ⅔(\mathcal{E}/V). \quad (83)$$

The specific heat at constant volume is readily found from the identity $C_V = \partial\mathcal{E}/\partial T$, where $\mathbb{N}$ and $V$ are kept constant. Thus, Eq.(81) yields $C_V = 3k_B\mathbb{N}/2$, which may equivalently be expressed as $C_V = 3R/2$ per mole. As for the specific heat at constant pressure, $C_\wp$, we note that raising the temperature by $\delta T$ at constant pressure $\wp$ raises the internal (kinetic) energy of the gas by $\delta\mathcal{E} = 3k_B\mathbb{N}\delta T/2$. At the same time, the volume $V$ expands by $\delta V$, which requires an additional work (i.e., input of energy) by $\wp\delta V = k_B T\mathbb{N}\delta V/V$; see Eq.(83). The constancy of $\wp = k_B\mathbb{N}T/V$ when both $T$ and $V$ vary while $\mathbb{N}$ remains fixed yields

$$\delta\wp = k_B\mathbb{N}[(\delta T/V) - (T\delta V/V^2)] = 0 \quad \rightarrow \quad \delta V/V = \delta T/T. \quad (84)$$

Thus, the total energy input when the temperature is raised by $\delta T$ while $\mathbb{N}$ and $\wp$ are kept fixed is

$$\delta\mathcal{E} + \wp\delta V = 3k_B\mathbb{N}\delta T/2 + k_B T\mathbb{N}\delta V/V = 5k_B\mathbb{N}\delta T/2 \quad \rightarrow \quad C_\wp = 5k_B\mathbb{N}/2. \quad (85)$$

The constancy of $\mathbb{N}$ is assured by setting $\delta\mathbb{N} = 0$ in Eq.(79), which yields

$$(3/2)T^{1/2}V\delta T + T^{3/2}\delta V - T^{3/2}V\delta\eta = 0 \;\rightarrow\; \delta\eta = (3/2)(\delta T/T) + (\delta V/V) = 5\delta T/2T. \quad (86)$$

Einstein proceeds to conclude that his "statistically-independent" distribution of distinguishable atoms among the available phase-space cells is incorrect, since the entropy of Eq.(82) — obtained from the distribution given by Eq.(70) — does *not* approach zero in the limit of $T \to 0$.[8] It seems that Einstein rushes to judgement in this instance, since the expression of $\ln(w)$ in Eq.(71) is inapplicable to the case of $T \to 0$. Nevertheless, the essence of his conclusion remains indisputable, since Bose's hypothesis as embodied by Eq.(34) — which is equivalent to Einstein's Eq.(63) for the number of arrangements of the particles among the phase-space cells — is far superior to the "statistically-independent" hypothesis pertaining to the phase-space distribution of atoms that forms the basis of Eq.(70).

**8. Concluding remarks**. In his letter of June 4, 1924 to Einstein, Bose wrote: "*I have ventured to send you the accompanying article for your perusal and opinion. I am anxious to know what you think of it. You will see that I have tried to deduce the coefficient $8\pi\nu^2/c^3$ in Planck's law independent of the classical electrodynamics, only assuming that the ultimate elementary regions in the phase-space have the content $h^3$.*" Einstein responded in a postcard dated July 2, 1924: "*I have translated your work and communicated it to Zeitschrift für Physik for publication. It signifies an important step forward and I liked it very much. Factually, I find your objections against my work not correct. For Wien's displacement law does not assume the wave (undulation) theory and Bohr's correspondence principle is not at all applicable. However, this does not matter. You are the first to derive the factor quantum theoretically, even though because of the polarization factor 2 not wholly rigorously. It is a beautiful step forward.*"[9]

The existence of the Bose-Einstein Condensate (BEC) was demonstrated for the first time in 1995 in the experiments of Eric Cornell, Carl Wieman, and Wolfgang Ketterle, who subsequently shared the 2001 Nobel Prize in Physics. The Nobel prize citation was "*for the achievement of Bose-Einstein condensation in dilute gases of alkali atoms, and for early fundamental studies of the properties of the condensates*."

There is an amusing story about Dirac's visit to Calcutta in 1954. Bose went with some of his students to the railway station to meet Dirac and his wife, who were taken to Bose's car and ushered into the back seat while Bose and his students crowded into the front seat. When Dirac, a founder of the Fermi-Dirac statistics, politely invited some students to come to the back seat, Bose quipped, "We believe in Bose statistics."[9]

## Appendix A

Consider two identical boxes of volume $V$, both in thermal equilibrium at temperature $T$, each of which contains a total of $\mathbb{N}$ identical particles. A fictitious membrane separating the boxes allows the particles in just one narrow energy range, say, $[\epsilon_s, \epsilon_s + \mathrm{d}\epsilon_s)$, to pass back and forth between the boxes; particles having any other energy remain confined within their own box. (Einstein likens this membrane to a narrow bandpass optical filter in case the particles were photons, which would then allow the exchange only of photons of a certain frequency, or color, between the boxes.) Let the number of particles crossing from box 1 to box 2 be denoted by $\delta M_s$. We will have

$$\delta \ln(w_{1,2}) \cong \frac{\partial \ln(w)}{\partial M_s}(\pm\delta M_s) + \frac{\partial^2 \ln(w)}{\partial M_s^2}(\pm\delta M_s)^2. \qquad \text{(A1)}$$

The entropies of the two boxes being additive under the circumstances, the overall change in the entropy of the two-box system, aside from the coefficient $k_B$, will be the sum of $\delta \ln(w)$ pertaining to boxes 1 and 2, namely,

$$\delta \ln(w_{1+2}) = \delta \ln(w_1) + \delta \ln(w_2) \cong 2\frac{\partial^2 \ln(w)}{\partial M_s^2}(\delta M_s)^2. \quad \text{(A2)}$$

The probability of a random exchange of $\delta M_s$ particles between boxes 1 and 2 must be proportional to the number of configurations $w_{1+2}$ that correspond to $\delta M_s$; that is,

$$\text{probability of } \delta M_s \propto \exp\left[2\frac{\partial^2 \ln(w)}{\partial M_s^2}(\delta M_s)^2\right]. \quad \text{(A3)}$$

Now, from Eq.(64), which is the same as Eq.(17), we find

$$\partial \ln(w)/\partial M_s = \ln(M_s + N_s - 1) + \cancel{1} - \ln(M_s) - \cancel{1}. \quad \text{(A4)}$$

Consequently,

$$\partial^2 \ln(w)/\partial M_s^2 = (M_s + N_s - 1)^{-1} - (M_s)^{-1} = -(N_s - 1)/[M_s(M_s + N_s - 1)]. \quad \text{(A5)}$$

Considering that the normal (or Gaussian) probability distribution function for a random variable $x$ with average $\bar{x}$ and standard deviation $\sigma_x$ is $(\sqrt{2\pi}\sigma_x)^{-1}\exp[-(x-\bar{x})^2/(2\sigma_x^2)]$, we conclude from Eqs.(A3) and (A5) that the variance of the particle-number fluctuations $\delta M_s$ is

$$\text{var}(M_s) = \frac{-1}{4\partial^2 \ln(w)/\partial M_s^2} = \frac{M_s(M_s + N_s - 1)}{4(N_s - 1)} \cong \frac{M_s^2}{4N_s} + \tfrac{1}{4}M_s. \quad \text{(A6)}$$

Normalizing the above variance of $M_s$ by the squared mean-value of the number of particles residing in the energy range $[\epsilon_s, \epsilon_s + \mathrm{d}\epsilon_s)$ inside each box, we finally arrive at

$$\frac{\text{var}(M_s)}{M_s^2} \cong \frac{1}{4M_s} + \frac{1}{4N_s}. \quad \text{(A7)}$$

Whereas the first term on the right-hand side of Eq.(A7) arises from the particle-number fluctuations, the second term, which is solely a property of the phase space, arises from interference effects similar to those of photons. Einstein knows this from his understanding of electromagnetic fluctuations in the case of photons and, therefore, associates the term $1/N_s$ of Eq.(A7) with Louis de Broglie's (then recent) hypothesis pertaining to the wave nature of material particles.

We close by pointing out that, in Einstein's original paper,[8] the second box is much larger than the first, albeit with the ratio $\mathbb{N}/V$ being the same for the two boxes. In that case, Eq.(A4) shows that the first derivative of $\ln(w)$ with respect to $M_s$ continues to be the same for the two boxes, whereas, in accordance with Eq.(A5), the second derivative of $\ln(w)$ would be negligible for box 2. Thus, in Einstein's original paper, Eq.(A2) appears without the factor of 2 on the right-hand side. The rest of the argument remains the same, of course, except for the final expression of the variance of $M_s$, which ends up being twice as large if box 2 is taken to be far larger than box 1.

## Appendix B

In arriving at Eq.(61), we used an approximate expression for $g_5(\eta)/g_3(\eta)$, which is valid for large values of $\eta$. A better approximation method, suggested in Einstein's second paper,[8] can be used to express $g_5(\eta)$ as a linear combination of the first few powers of $g_3(\eta)$. The method involves setting $e^{-\eta} = x$, then writing $g_5(\eta)$ as $F(x) = \sum_{n=1}^{\infty} n^{-5/2}x^n$ and $g_3(\eta)$ as $G(x) = \sum_{n=1}^{\infty} n^{-3/2}x^n$. Given that $F(x)$ and $G(x)$ are well-defined and uniformly increasing functions over $0 \le x \le 1$, and that $F(0) = G(0) = 0$, one can take $F$ to be a well-behaved function of $G$, and proceed to expand $F$ in a Taylor series around $G = 0$, as follows:

$$F(x) = \tilde{F}[G(x)] = \frac{\mathrm{d}\tilde{F}}{\mathrm{d}G}\bigg|_{x=0} G(x) + \frac{1}{2!}\frac{\mathrm{d}^2\tilde{F}}{\mathrm{d}G^2}\bigg|_{x=0} G^2(x) + \frac{1}{3!}\frac{\mathrm{d}^3\tilde{F}}{\mathrm{d}G^3}\bigg|_{x=0} G^3(x) + \frac{1}{4!}\frac{\mathrm{d}^4\tilde{F}}{\mathrm{d}G^4}\bigg|_{x=0} G^4(x) + \cdots. \quad \text{(B1)}$$

The first few derivatives of $\tilde{F}(G)$ with respect to $G$ may now be computed iteratively, as follows:

$$\frac{\mathrm{d}\tilde{F}}{\mathrm{d}G}\bigg|_{x=0} = \frac{\mathrm{d}F/\mathrm{d}x}{\mathrm{d}G/\mathrm{d}x}\bigg|_{x=0} = \left(\sum_{n=1}^{\infty} n^{-3/2}x^{n-1} \Big/ \sum_{n=1}^{\infty} n^{-1/2}x^{n-1}\right)_{x=0} = 1. \quad \text{(B2)}$$

$$\frac{\mathrm{d}^2\tilde{F}}{\mathrm{d}G^2}\bigg|_{x=0} = \frac{\mathrm{d}(\mathrm{d}\tilde{F}/\mathrm{d}G)/\mathrm{d}x}{\mathrm{d}G/\mathrm{d}x}\bigg|_{x=0} = \left[\frac{\sum_{n=1}^{\infty}(n-1)n^{-3/2}x^{n-2}}{(\sum_{n=1}^{\infty} n^{-1/2}x^{n-1})^2} - \frac{\sum_{n=1}^{\infty} n^{-3/2}x^{n-1}\sum_{n=1}^{\infty}(n-1)n^{-1/2}x^{n-2}}{(\sum_{n=1}^{\infty} n^{-1/2}x^{n-1})^3}\right]_{x=0}$$

$$= \frac{1}{2^{3/2}} - \frac{1}{2^{1/2}} = -\frac{1}{2\sqrt{2}} \cong -0.3536. \quad \text{(B3)}$$

$$\frac{\mathrm{d}^3\tilde{F}}{\mathrm{d}G^3}\bigg|_{x=0} = \frac{\mathrm{d}(\mathrm{d}^2\tilde{F}/\mathrm{d}G^2)/\mathrm{d}x}{\mathrm{d}G/\mathrm{d}x}\bigg|_{x=0}$$

$$= \left[\frac{\sum_{n=1}^{\infty}(n-1)(n-2)n^{-3/2}x^{n-3}}{(\sum_{n=1}^{\infty} n^{-1/2}x^{n-1})^3} - \frac{3\sum_{n=1}^{\infty}(n-1)n^{-3/2}x^{n-2}\sum_{n=1}^{\infty}(n-1)n^{-1/2}x^{n-2}}{(\sum_{n=1}^{\infty} n^{-1/2}x^{n-1})^4}\right.$$

$$\left. - \frac{\sum_{n=1}^{\infty} n^{-3/2}x^{n-1}\sum_{n=1}^{\infty}(n-1)(n-2)n^{-1/2}x^{n-3}}{(\sum_{n=1}^{\infty} n^{-1/2}x^{n-1})^4} + \frac{3\sum_{n=1}^{\infty} n^{-3/2}x^{n-1}[\sum_{n=1}^{\infty}(n-1)n^{-1/2}x^{n-2}]^2}{(\sum_{n=1}^{\infty} n^{-1/2}x^{n-1})^5}\right]_{x=0}$$

$$= \frac{2}{3^{3/2}} - \frac{3}{2^{3/2}\cdot 2^{1/2}} - \frac{2}{3^{1/2}} + \frac{3}{2} \cong -0.0198. \quad \text{(B4)}$$

$$\frac{\mathrm{d}^4\tilde{F}}{\mathrm{d}G^4}\bigg|_{x=0} = \frac{\mathrm{d}(\mathrm{d}^3\tilde{F}/\mathrm{d}G^3)/\mathrm{d}x}{\mathrm{d}G/\mathrm{d}x}\bigg|_{x=0} = \left[\frac{\sum_{n=1}^{\infty}(n-1)(n-2)(n-3)n^{-3/2}x^{n-4}}{(\sum_{n=1}^{\infty} n^{-1/2}x^{n-1})^4}\right.$$

$$- \frac{6\sum_{n=1}^{\infty}(n-1)(n-2)n^{-3/2}x^{n-3}\sum_{n=1}^{\infty}(n-1)n^{-1/2}x^{n-2}}{(\sum_{n=1}^{\infty} n^{-1/2}x^{n-1})^5}$$

$$- \frac{4\sum_{n=1}^{\infty}(n-1)n^{-3/2}x^{n-2}\sum_{n=1}^{\infty}(n-1)(n-2)n^{-1/2}x^{n-3}}{(\sum_{n=1}^{\infty} n^{-1/2}x^{n-1})^5}$$

$$+ \frac{15\sum_{n=1}^{\infty}(n-1)n^{-3/2}x^{n-2}[\sum_{n=1}^{\infty}(n-1)n^{-1/2}x^{n-2}]^2}{(\sum_{n=1}^{\infty} n^{-1/2}x^{n-1})^6}$$

$$- \frac{\sum_{n=1}^{\infty} n^{-3/2}x^{n-1}\sum_{n=1}^{\infty}(n-1)(n-2)(n-3)n^{-1/2}x^{n-4}}{(\sum_{n=1}^{\infty} n^{-1/2}x^{n-1})^5}$$

$$+ \frac{10\sum_{n=1}^{\infty} n^{-3/2}x^{n-1}\sum_{n=1}^{\infty}(n-1)(n-2)n^{-1/2}x^{n-3}\sum_{n=1}^{\infty}(n-1)n^{-1/2}x^{n-2}}{(\sum_{n=1}^{\infty} n^{-1/2}x^{n-1})^6}$$

$$\left. - \frac{15\sum_{n=1}^{\infty} n^{-3/2}x^{n-1}[\sum_{n=1}^{\infty}(n-1)n^{-1/2}x^{n-2}]^3}{(\sum_{n=1}^{\infty} n^{-1/2}x^{n-1})^7}\right]_{x=0}$$

$$= \frac{6}{4^{3/2}} - \frac{12}{3^{3/2}\cdot 2^{1/2}} - \frac{8}{2^{3/2}\cdot 3^{1/2}} + \frac{15}{2^{3/2}\cdot 2} - \frac{6}{4^{1/2}} + \frac{20}{3^{1/2}\cdot 2^{1/2}} - \frac{15}{2^{3/2}} \cong -0.00267. \quad \text{(B5)}$$

Consequently,

$$F(x) \cong G(x) - 0.1768\, G^2(x) - 0.0033\, G^3(x) - 0.00011\, G^4(x) + \cdots. \quad \text{(B6)}$$

This is nearly the same expansion of $F(x)$ in powers of $G(x)$ as given in Einstein's second paper,[8] the main difference being the coefficient of $G^4(x)$, which Einstein specifies as $-0.0005$.

---

The following identities, listed in the *Handbook of Integrals, Series, and Products*[4], have been used throughout this article:

G&R **3.411**-1: $\int_0^\infty \frac{x^{\nu-1}}{e^{\mu x}-1}\mathrm{d}x = \Gamma(\nu)\zeta(\nu)/\mu^\nu,$ [$\zeta(\cdot)$ is Riemann's zeta function.] (B7)

G&R **9.542**-1, **9.620**, and **9.627**-4: $\zeta(4) = \sum_{n=1}^{\infty} n^{-4} = \pi^4/90.$ (B8)

G&R **4.262**-2: $\int_0^1 \frac{(\ln x)^3}{1-x}\mathrm{d}x = -\frac{\pi^4}{15};$ change of variable: $x = e^{-y} \rightarrow \int_0^\infty \frac{y^3}{e^y - 1}\mathrm{d}y = \frac{\pi^4}{15}.$ (B9)

G&R **3.351**-3: $\int_0^\infty x^n e^{-\mu x}\mathrm{d}x = n!\,\mu^{-(n+1)},$ [$\mathrm{Re}(\mu) > 0$]. (B10)

G&R **3.381**-4: $\int_0^\infty x^{\nu-1} e^{-\mu x}\mathrm{d}x = \Gamma(\nu)/\mu^\nu,$ [$\mathrm{Re}(\mu) > 0$, $\mathrm{Re}(\nu) > 0$]. (B11)

$$\Gamma(x+1) = x\Gamma(x), \qquad \Gamma(1) = 1, \qquad \Gamma(½) = \sqrt{\pi}. \tag{B12}$$